\documentclass[lettersize,journal]{IEEEtran}
\usepackage{amsmath,amsfonts}
\usepackage{algorithmic}
\usepackage{array}
\usepackage[caption=false,font=normalsize,labelfont=sf,textfont=sf]{subfig}
\usepackage{textcomp}
\usepackage{stfloats}
\usepackage{url}
\usepackage{verbatim}
\usepackage{float}
\usepackage{xcolor,soul,framed} 
 \usepackage[subtle,tracking=normal]{savetrees}
\colorlet{shadecolor}{yellow}
\usepackage[pdftex]{graphicx}

\usepackage{cite}
\usepackage{nomencl} 
\makenomenclature 
\renewcommand\nomgroup[1]{%
\ifthenelse{\equal{#1}{M}}{\item[\textbf{Sets and Indices}]}{%
 \ifthenelse{\equal{#1}{O}}{\item[\textbf{Decision variables for DA problem}]}{}}{%
 \ifthenelse{\equal{#1}{N}}{\item[\textbf{Day-ahead Parameters}]}{}}{%
  \ifthenelse{\equal{#1}{T}}{\item[\textbf{Real-time Parameters}]}{}}{%
   \ifthenelse{\equal{#1}{W}}{\item[\textbf{Real-time Outputs}]}{}}{%
    \ifthenelse{\equal{#1}{P}}{\item[\textbf{Dual decision variables}]}{}}%

}

\begin{document}

\title{Addressing Imbalance Risk with Reserves and Flexibility Options: An ERCOT-like Case Study}

\author{Elina Spyrou, Robin Hytowitz, Benjamin F. Hobbs, Ibrahim Krad, Liping Li, Mengmeng Cai, Michael Blonsky
\thanks{{This work was supported in part by funding provided by the U.S. Department of Energy Advanced Research Projects Agency–Energy (ARPA-E) under ARPA-E Award No. DE-AR001274 and Work Authorization No. 19/CJ000/07/08 and in part by the Leverhulme Trust. }}
}

\markboth{This work has been submitted to the IEEE for possible publication. Copyright may be transferred without notice, after which this version may no longer be accessible.}{}


\maketitle

\begin{abstract}
As the role of variable renewables in electricity markets expands, new market products help system operators manage imbalances caused by uncertainty and variability. Whereas work in the last decade has focused on constructing demand curves for central procurement of those products, little attention has been paid to designing their settlement scheme and understanding the connections between the economic value of these products, the schedule of variable resources, and the cost of flexibility. In this article, we compare a new product called Flexibility Options, which addresses these gaps, with a traditional reserve product using a case study similar to the 2019 Texas (ERCOT) system. Our findings suggest that both products are equally effective in managing imbalances, but Flexibility Options have superior risk management properties and keep the system operator revenue adequate.
\end{abstract}


\newcommand\myuparrow{\mathord{\uparrow}}	

\newcommand{\hsup}{{hs}^{\uparrow}}
\newcommand{\hsdn}{{hs}^{\downarrow} }
\newcommand{\hdup}{{hd}^{\uparrow}}
\newcommand{\hddn}{{hd}^{\downarrow}}
\newcommand{\hdupdn}{{hd}^{\uparrow|\downarrow}}
\newcommand{\hsupdn}{{hs}^{\uparrow|\downarrow}}

\newcommand{\regup}{{reg}^{\uparrow}}
\newcommand{\regdn}{{reg}^{\downarrow}}
\newcommand{\regupdn}{{reg}^{\uparrow|\downarrow}}

\newcommand{\sdu}{{sd}^{\uparrow}}
\newcommand{\sdd}{{sd}^{\downarrow}}
\newcommand{\sdud}{{sd}^{\uparrow|\downarrow}}

\newcommand{\lda}{\lambda^{DA}}
\newcommand{\lrt}{\lambda^{RT}}
\newcommand{\lup}{{\lambda}^{\uparrow}}
\newcommand{\ldn}{{\lambda}^{\downarrow}}
\newcommand{\lupdn}{{\lambda}^{\uparrow|\downarrow}}

\newcommand{\capup}{{CAP}^{\uparrow}}
\newcommand{\capdn}{{CAP}^{\downarrow}}
\newcommand{\capupdn}{{CAP}^{\uparrow|\downarrow}}

\newcommand{\ccup}{C^{\uparrow}}
\newcommand{\ccdn}{C^{\downarrow}}
\newcommand{\ccupdn}{C^{\uparrow|\downarrow}}

\newcommand{\prup}{{\Pi}^{\uparrow}}
\newcommand{\prdn}{{\Pi}^{\downarrow}}
\newcommand{\prupdn}{{\Pi}^{\uparrow|\downarrow}}

\newcommand{\Pmin}{P^{min}}
\newcommand{\Pmax}{P^{max}}

\newcommand{\RTHSup}{\overline{HS}^{\uparrow}}
\newcommand{\RTHSdn}{\overline{HS}^{\downarrow}}
\newcommand{\RTHSupdn}{\overline{HS}^{\uparrow|\downarrow}}
\newcommand{\kup}{\alpha^{\uparrow}}
\newcommand{\kdn}{\alpha^{\downarrow}}
\newcommand{\kupdn}{\alpha^{\uparrow|\downarrow}}
\newcommand{\RTHDup}{\overline{HD}^{\uparrow}}
\newcommand{\RTHDdn}{\overline{HD}^{\downarrow}}
\newcommand{\RTHDupdn}{\overline{HD}^{\uparrow|\downarrow}}

\newcommand{\vcup}{{VC}^{\uparrow}}
\newcommand{\vcdn}{{VC}^{\downarrow}}
\newcommand{\vcupdn}{{VC}^{\uparrow|\downarrow}}

\newcommand{\reupda}{{RE}^{DA,\uparrow}}
\newcommand{\rednda}{{RE}^{DA,\downarrow}}
\newcommand{\reupdnda}{{RE}^{DA,\uparrow|\downarrow}}
\newcommand{\reuprt}{{RE}^{RT,\uparrow}}
\newcommand{\rednrt}{{RE}^{RT,\downarrow}}
\newcommand{\reupdnrt}{{RE}^{RT,\uparrow|\downarrow}}

\newcommand{\payupda}{{PAY}^{DA,\uparrow}}
\newcommand{\paydnda}{{PAY}^{DA,\downarrow}}
\newcommand{\payupdnda}{{PAY}^{DA,\uparrow|\downarrow}}
\newcommand{\payuprt}{{PAY}^{RT,\uparrow}}
\newcommand{\paydnrt}{{PAY}^{RT,\downarrow}}
\newcommand{\payupdnrt}{{PAY}^{RT,\uparrow|\downarrow}}
\newcommand{\ccrt}{C^{RT}} 
 \newcommand{\og}{\Omega_G}
 \newcommand{\ogs}{\Omega_{G}^{S}}
 \newcommand{\ogb}{\Omega_{G}^B}
\newcommand{\HSup}{{HS}^{\uparrow}}
\newcommand{\HSdn}{{HS}^{\downarrow} }
\newcommand{\HSupdn}{{HS}^{\uparrow|\downarrow} }
\newcommand{\HDup}{{HD}^{\uparrow}}
\newcommand{\HDdn}{{HD}^{\downarrow}}
\newcommand{\HDupdn}{{HD}^{\uparrow|\downarrow}}

\newlength\myindent
\setlength\myindent{0em}
\newcommand\bindent{%
  \begingroup
  \setlength{\itemindent}{\myindent}
  \addtolength{\algorithmicindent}{\myindent}
}
\newcommand\eindent{\endgroup}

\newcolumntype{E}{>{\centering\arraybackslash}m{1cm}}
\newcommand{\R}{\mathbb{R}}
\nomenclature[M]{$AS$}{Set of ancillary services, indexed $a$}
\nomenclature[M]{${AS}^{\uparrow|\downarrow}$}{Subset of ancillary services in up or down direction}
\nomenclature[M]{$G$}{Set of market participants, indexed $i$}
\nomenclature[M]{$G^{B}$}{Set of FO buyers, subset of $G$ }
\nomenclature[M]{$G^{S}$}{Set of FO sellers, subset of $G$ }
\nomenclature[M]{${T}$}{Set of time intervals, indexed $t$}
\nomenclature[M]{${S}$}{ Set of scenarios, indexed $s$, in ascending order wrt electricity generation. That is, $s =1$ corresponds to the lowest electricity generation of a FO buyer. }
\nomenclature[M]{${R}$}{ Set of tiers, indexed $r$. $R = \{ 1, \ldots, r, \ldots, \lvert S\rvert -1 \} $}

\nomenclature[O]{$\hsupdn_{i,r,t}$}{FO up/down at tier $r$, time $t$ sold by $i\in G^S$(MW) }
\nomenclature[O]{$\hdupdn_{i,r,t}$}{FO up/down at tier $r$, time $t$ bought by $i \in G^B$(MW)} 
\nomenclature[O]{$p^{DA/RT}_{i,t}$ }{Participant's $i$ DA/RT electricity sales at $t$ (in MWh) } 
\nomenclature[O]{$\sdud_{i,r,t}$ }{ Self-hedged up/down imbalance at tier $r$ for buyer $i \in G^B$  (MW) }
\nomenclature[O]{$y_{i,s,t}$ }{ Auxiliary variable for absolute FO volume hedging FO buyer $i \in G^B$, in scenario $s$ at $t$ (MW)}
\nomenclature[O]{$u_{i,t}$ }{ Unit status of generator $i$ at $t$; $ u_{i,t} \in \{0,1\}$ }
\nomenclature[O]{${res}_{i,a,t}$}{Reserve award for service $a$, time $t$ sold by $i\in G(MW)$ }
\nomenclature[P]{$\lambda^{DA/RT}_t$}{DA/RT energy price at interval $t$ (\$/ MWh)}
\nomenclature[P]{$\lupdn_{r,t}$ }{DA FO up/down price of tier $r$ at time $t$ (in \$/ MW) }

\nomenclature[N]{$\widetilde{\vcupdn_{i,t}}$}{Up/down FO scarcity cost for  FO buyer $i \in G^B$ at  $t$ } 
\nomenclature[N]{$C_{i,t}$ }{ Variable cost for participant $i$ at time $t$ (\$ / MWh) }
\nomenclature[N]{$\beta_a$}{configurable parameter that reserves ramping capability for reserve $a$}
\nomenclature[N]{$\ccupdn_{i,t}$ }{RT strike price for FO up/down by seller $i \in G^S$ at time $t$ (\$/MWh)  }
\nomenclature[N]{$M$ }{Penalty term for $y_{i,s,t}$ ($\$/MW$) }
\nomenclature[N]{$\Pi_{s}$ }{Probability of scenario $s$  }
\nomenclature[N]{$\prupdn_{r}$ }{ Probability of exercising FO up/down of tier $r$    }
\nomenclature[N]{$P_{i,s,t}$ }{ RT upper operating limit for FO buyer $i \in G^B$ at  $t$ in scenario $s$  (MWh)}
\nomenclature[N]{$\Pmin_{i,t}$ }{ Minimum generation when participant $i$ on at $t$ (MW) }
\nomenclature[N]{$\Pmax_{i,t}$ }{ Generator capacity of participant $i$ at $t$ (MW)}
\nomenclature[N]{$RR_{i,t}$ }{ Ramp rate of FO supplier $i$ at $t$ (MW/hour) }

\printnomenclature

\section{Introduction}
\IEEEPARstart{P}{ower} system operators globally are developing new solutions and services to operate systems that integrate large amounts of variable renewable resources \cite{FERC2021, Balarko_services_beyond_energy}. These services aim to address operability challenges cost-effectively and reliably. In a sequential market setup, one set of such services \textit{reserve} capacity and sometimes \textit{energy} in a forward market (e.g., Day-Ahead (DA)) and release the reserved capacity in a subsequent market (e.g., Real-time (RT)). 

To fully define a service, system operators have to make several design choices. Ref. \cite{Glismann_Nobel} lists fourteen such design choices. The choices specify how and by whom \textit{demand} will be determined; who is eligible to \textit{supply} the product; what processes will be followed for \textit{procurement and activation}; what  \textit{pricing and financial settlement scheme} will be in place to reward suppliers for their services and to allocate product costs to market participants. For each choice, several alternatives exist, and insights from operating experience, experiments, or the literature can shed light on their relative strengths and weaknesses. Reviewing the literature, we find that some design choices have been extensively discussed, while others have received less attention.  Interestingly, some topics that were extensively studied in the early years of industry restructuring \cite{ISEMONGER2009150} are being revisited as the electricity sector transitions to a future of low greenhouse gas emissions. One such topic concerns the benefits of co-optimizing energy and reserves in continental Europe and Great Britain \cite{Smeers_cooptimization, Pap_ACER, ENA_coopt}.

Several papers focus on construction of demand curves for central procurement and valuation of operating reserves. While most of them concentrate on the horizontal axis of the demand curve (i.e., sizing), a few papers also discuss the vertical axis of the demand curve (i.e., scarcity pricing). 
Regarding sizing, as \cite{DEVOS2019272} points out, two types of methods have been proposed: \textit{bottom-up} methods which minimize probability-weighted costs and penalties for unserved demand; and \textit{probabilistic methods} which aim to keep the probability of a RT capacity shortfall below a certain target. 

Early on, researchers and practitioners recognized that the demand for these products should be dynamic because reserve needs depend on weather \cite{milligan2010operating}. Since then, numerous articles have proposed and tested various data-driven methods for dynamic sizing of reserve products \cite{DEVOS2019272, Binghui_SF2, EPRI_DOE_Forecasting,LAVIN2020111857, E3_PERFORM}. The literature has repeatedly shown that dynamic requirements outperform static ones in two ways: by saving costs on some days, while improving reliability on other days. Whereas sizing is typically studied at the system level, industry reports indicate that reserved capacity may be sometimes undeliverable due to transmission constraints \cite{CAISO_FRP_report}. To address this issue, some power system operators apply post-reserve deployment transmission constraints and quantify uncertainties at the system, sub-regional, or nodal levels \cite{Yongchong_MISO, FRP_deployment_Scenarios, Nodal_vs_zonal_Deployment_constraints}. Academics have also explored methods such as dynamic partitioning of the system in reserve zones \cite{Dynamic_partitioning} and generator-specific reserve response factors \cite{Nikita_reserve_response} as potential solutions to deliverability problems.

Most methods on reserve sizing make the strong assumption that the DA schedule of variable resources will match the day-ahead `deterministic forecast', overlooking that variable resources might follow forward strategies for managing forecast errors that result in day-ahead schedules deviating from deterministic forecasts \cite{Dent_Bialek_Hobbs}. The forecast provides a reference point for estimating RT imbalances in the upward and downward directions. For accurate estimation of RT imbalances, the reference point  should be the DA net load schedule. However, the system operator must determine reserve requirements before solving the DA market, which provides the DA net load schedule. In the design of Imbalance Reserves (IR)\footnote{A DA reserve product to manage DA-to-RT net load imbalances.}, the CAISO acknowledged this issue and suggested introducing a complementary product in a subsequent reliability process to correct any deviation between the DA net load forecast and schedule.  A method involving solution of sequential optimization problems together with submission of uncertainty bids by market participants (i.e., minimum and maximum production levels) has also been proposed in the literature \cite{SILVARODRIGUEZ2024121982}. However, optimal solutions to sequential problems do not necessarily result in overall optimal decisions. Another article  presents a truly endogenous reserve sizing method within the DA market \cite{Endogenous}. This method gives more accurate estimates of the need for and value of reserves, and can schedule reserves from variable resources, thereby considering the trade-off between variable energy schedules and demand for reserves. However, \cite{Endogenous} only partially addressed the need to link reserve needs with DA net load schedules, as it does not discuss reserve pricing and settlement schemes.

Turning to the actual practice of reserve scarcity pricing, operator practices in the USA vary widely \cite{scarcity_price_usa}. In theory, when reserves are released in a subsequent market with uniform pricing, the reserve scarcity price should reflect the probability-weighted value of lost load\footnote{The value of lost load is appropriate for systems that have a single operating reserve product. However, in systems where there are multiple operating reserves, the system might lean on other types of reserves when one type is missing. In those cases, instead of the value of lost load, the scarcity price of the reserves (e.g., contingency reserves) the system leans on when the reserve in question is missing should be used. } minus the probability-weighted energy price of the subsequent market \cite{Hogan_ORDC_2013}. To estimate reserve scarcity prices, \cite{Cartuyvels_calibration} and \cite{mays2021quasi} rely on Monte Carlo simulations of nested optimization problems and stochastic programs, respectively. Both of those papers implicitly assume that the system operator knows in advance (e.g., in DA) the RT costs of reserve suppliers, concurring with \cite{ELA201651, Endogenous} that the cost of providing flexibility matters. In practice, while designing the IR product, CAISO was however unable to convince stakeholders that it could even estimate an upper bound to these costs \cite{DAME_DraftFinal, DAME_EligibilityCap, DAME_stakeholders, DAME_Final}, much less their expected value. This striking divergence between industry experience and academic literature suggests that design aspects with respect to flexibility costs merit further investigation.

Similarly, settlement schemes that allocate costs of reserves to participants have not been extensively discussed. Whereas this might seem surprising at first, it might be explainable by the historically small size of reserve markets \cite{EPRI_AStrends}. Ref. \cite{Kirschen_2021} argues that by charging parties contributing to the outage risk the costs of mitigation measures (e.g., reserves), these parties are financially incentivized to reduce this risk and the associated cost. This principle underlies proposals by \cite{Carlos_TEMPR2024} and \cite{Haring_Kirschen_Andersson}, who present cost-allocation schemes for contingency and non-contingency reserves, respectively. Both \cite{Carlos_TEMPR2024} and \cite{Haring_Kirschen_Andersson} are computationally expensive as they solve a series of optimisation problems in order to allocate the costs. 

Whereas most reserve settlement schemes do not fully allocate costs to parties contributing to the risk, the settlement scheme for the IR product in CAISO charges parties with observed imbalances \cite{DAME_Final}. This is an ex-post cost-causation link because it does not attempt to estimate each party's contribution to the need of reserves ex-ante (i.e., in DA) when reserve requirements are often determined. Ex-post causation links that compare actual resource output or load to an expected trajectory have been used in Australia for many years \cite{RIESZ201586}. However, under uncertainty, ex-post cost-causation links might not fully recover the reserve costs or provide inefficient incentives. 

In summary, we identify at least three design issues for reserve products that warrant further investigation: (1) the link between the demand for reserves and the DA net load schedule; (2) ex-ante assumptions on the balancing costs of reserve suppliers; and (3) settlement schemes. To address these issues, we proposed a product in \cite{FOpaper1} that we call Flexibility Options (FO). FO establish a link between the demand for reserves and the DA net load schedule by allowing participants susceptible to imbalances to buy FO that hedge against multiple possible RT output levels with different probabilities of a shortfall (or a surplus). The formulation uses those levels and the probabilities to co-optimize the demand for reserves and the day-ahead energy schedule of resources with imbalances. To form assumptions on balancing costs, FO financial settlements adhere to the principles of a dual-trigger option. Dual-trigger products are `policies that are activated in the event of two contingencies' \cite{COX2004259}.  One of the conditions for FO activation (i.e., trigger) is the familiar strike price for exercising the option (giving the holder the right to put or call a MW quantity), in which suppliers financially commit to strike prices in DA. These strike prices are the DA market inputs on the RT balancing costs of flexible units, which the system operator uses to price the product in DA.  The second condition for FO activation is the output of the risky resource or an aggregate measure (e.g., net load) exceeding or falling short with respect to a trigger quantity. The dual-trigger format of the product enhances risk management for both suppliers and buyers. For instance, by making the price protection provided by the call or put strike price available only if a resource's RT imbalance exceeds a certain amount, the payoff of the instrument is focused on circumstances when the resource most needs the insurance against financial losses from imbalances. To establish technically sound cost-causation links that keep the system operator revenue neutral, the settlement scheme follows standard option pricing with two components: a premium in DA and a payoff in RT. 

The FO product's properties have been explained via simple examples and mathematical analysis in \cite{FOpaper1}. In the present article, we provide evidence from a realistic case study to help system operators appreciate the significance of these three design issues and show the value of FO. 

Our case study has several attributes that make it valuable for informing discussions on the design of products that address imbalances caused by forecast errors. First, it relies on sequential optimization problems to solve DA and RT markets, reflecting the multi-settlement structure of actual markets. Second, we use scenarios that characterize net load uncertainty in DA, which were generated by cutting-edge methods\cite{ORFEUS_data} . Additionally, we use synthetic actual data (out-of-sample) to conduct the RT simulations. By comparing simulation results with FO vs. with a more traditional reserve product (similar to IR), we quantify the relative impact of FO on both system cost and the net revenues of market participants, including both flexible participants and participants susceptible to imbalances due to forecast errors.

We find that in this case, the two products are almost equally effective in managing imbalances, as the average system costs with the two products are quite similar. However, the FO result in less steep RT cost curves, benefiting both flexible suppliers and buyers by reducing the volatility of flexibility-related payments. Furthermore, our analysis demonstrates that unlike an Imbalance Settlement (IR) design with ex-post causation links that only recovers about 50\% of reserve costs, the FO settlement scheme consistently recovers 100\%, keeping the system operator revenue neutral. 

The rest of this paper is structured as follows. Section II provides background on the design of the two products. Section III presents the simulation environment. Section IV introduces the case study, whose results are presented in Section V. Last, Section VI summarizes our conclusions.

\section{Background: IR and FO Products}
We here provide a brief description of the two products, starting with their common design choices and concluding this section with a summary of their differences.  

Both IR and FO products reserve capacity in the upward and downward direction to manage DA-to-RT net load forecast errors. They are procured at the system level in the DA market. Their procurement is co-optimized within the DA market, and as a result, the product awards have the same resolution (hourly) and horizon (24 hours) as the rest of the DA market products. The pool of eligible suppliers includes on-line units with headroom and ramping capability in the up or down directions, as well as off-line units with start-up times, minimum up and down times small enough to be started within the lead time of the RT unit commitment and to have their commitment status adjusted within the horizon of the RT unit commitment (3 hours).  

The reserved capacity from both products is released in the RT market, which operates with a 15-minute resolution. In the case of IR, suppliers must submit RT bids that cover outputs with a range consistent with their IR awards. In the case of FO, suppliers are obligated to pay the payoffs when the FO are activated (i.e., satisfy the two triggers\footnote{We have assumed automatic exercise of FO when the two conditions for the triggers are met, but future research could investigate if this is an optimal strategy.}).

The demand for the two products is determined differently. For IR, the system operator constructs a reserve demand curve. The reserves are estimated based on historical or projected distributions of deviations from the DA net load forecast, with the marginal value of reserves declining as more are procured, acknowledging that holding more reserves reduces the probability of a real-time capacity shortfall but has diminishing marginal returns. In contrast, the demand for FO is endogenously determined based on possible deviations from the DA net load schedule and submitted projections for the RT net load components.  The FO prices are also endogenously determined by incorporating the probability-weighted strike prices that FO suppliers commit to in the DA objective function. 

To introduce IR/FO in DA markets, new information is needed. For IR, the system operator needs to generate the demand curve. For FO, the system operator needs to select probability levels to consider, the FO suppliers need to submit strike-up and -down prices, and the FO buyers need to submit levels for their projected output that align with the system operator's chosen probability levels. In both schemes, suppliers could also submit reserve capacity bids (i.e., costs that are incurred in DA and depend on the level of reserved capacity).

Both settlement schemes have DA and RT components. In the case of IR, the DA component rewards IR suppliers. In the case of FO, the DA component rewards FO suppliers and charges FO buyers. As for RT, the IR scheme charges participants with imbalances some or all of the DA IR costs; whereas in the FO scheme, when the options are exercised, the FO buyers receive option payoffs that the FO sellers pay.

\section{Methodology}
\begin{figure}[htb!]
\small
\begin{center}
\includegraphics[scale=1]{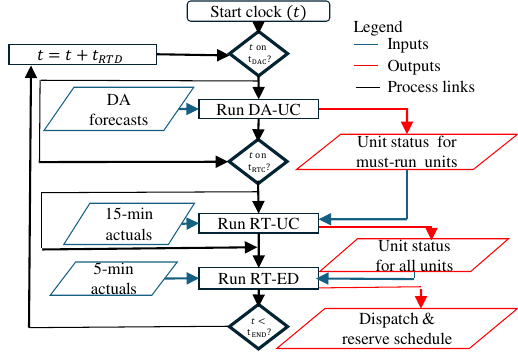} 
\caption{FESTIV flowchart in this study.}
    \label{fig:FESTIV}
\end{center}
\end{figure}

We simulate short-term power system operations using the Flexible Energy Scheduling Tool for
Integration of Variable generation (FESTIV) \cite{ELA_FESTIV}. We choose FESTIV because it has a sequential model structure including DA Unit Commitment, hour-ahead RT unit Commitment (RTC), and RT economic Dispatch (RTD) (see Fig. \ref{fig:FESTIV}).\footnote{FESTIV has a module that simulates Load Frequency Control, commonly known as AGC. In this application, we have skipped this module as it is not essential for the analysis, and omitting it reduces the computational time.} We include the two products in question (FO, IR) in the DA-UC and rely on the RT modules for managing the imbalances caused by load, wind, and solar forecast errors. In the RT modules, we release all capacity that was reserved for FO/IR in DA. We also allow for re-dispatch of all units and re-commitment of units without fixed status. This choice makes our implementation generous as the system can lean in RT on flexibility that was not necessarily contracted through IR or FO in DA. 

We model the IR product using the readily-available template for reserve products in FESTIV. In particular, we model IR up as a set of reserve products with different levels of scarcity prices, response time 60 minutes, with eligible suppliers including offline units that can start within 60 minutes and online units. To get a single IR up price, we use the FESTIV feature of cascading reserves allowing reserves with higher scarcity prices to contribute to reserves with lower scarcity prices. We modified the standard FESTIV template only with respect to inter-temporal ramping constraints. Our modification ensured that the ramping constraints would be respected even when a unit moves from fully deploying down reserves to fully deploying up reserves within an hour and vice versa. For IR down, we follow a similar approach.

For the FO, we modify the DA formulation by including additional constraints, modifying a few ramping constraints and the objective function. Below is a brief summary of the equations we added in FESTIV.

\begin{equation}
\small
\label{optprobDA}
		\begin{aligned}
		&\underset{\Xi}{\text{minimize}} \hspace{12pt} && {Objective}^{DA}  \\
& \text{ where } & &\Xi = \{ p^{DA}_{i,t}, {res}_{res,i,t},\hsupdn_{i,r,t}, 
\hdupdn_{i,r,t}
\sdud_{i,r,t},  u_{i,t}, y_{i,s,t}\} 
		\end{aligned}
	\end{equation}
\begin{subequations}\
\small
The additional terms in the ${Objective}^{DA}$ are given below:	
\begin{equation}\label{eqn_1b}
\begin{split}
	 {Additional}^{DA} =  \
	\pm \sum\nolimits_{i\in G^S, r \in R, t \in T} \ \prupdn_{r,t} \cdot \ccupdn_{i,t} \cdot \hsupdn_{i,r,t}\
\end{split}
\end{equation}  
\begin{equation}\label{eqn_1c}
	\begin{split}
		\mp \sum\nolimits_{i\in G^B, r \in R, t \in T}   \left(\prupdn_{r,t} \cdot C_{i,t} \cdot (\hdupdn_{i,r,t}+\sdud_{i,r,t})\right) \
	\end{split}
\end{equation}  
\begin{equation}\label{eqn_1d}
	\begin{split}
		\pm \sum\nolimits_{i\in G^B, r \in R, t \in T}  \ \left(\prupdn_{r,t} \cdot \widetilde{\vcupdn_{i,t}} \cdot \sdud_{i,r,t}\right) \\ 
	\end{split}
\end{equation} 
\begin{equation}\label{eqn_1e}
	+ M \cdot \sum_{i, s, t} y_{i,s,t}
\end{equation}
\end{subequations}
Note that $\pm$ above indicate that there is one term for FO in the up direction that is added, and one term for FO down that is subtracted ($\mp$ indicates the reverse).
The above objective is minimized subject to the following \textit{additional DA} constraints:
\begin{center}
\small
$p^{DA}_{i,t}, \hsupdn_{i,r,t}, \hdupdn_{i,r,t} ,\sdud_{i,r,t}, y_{i,s,t} \geq 0$ \
\end{center}
\begin{equation}\label{eqn_4}
\sum\limits_{i\in G^S} \hsupdn_{i,r,t} = \sum\limits_{i\in G^B} \hdupdn_{i,r,t}  \hspace{10pt} 
\forall r \in R,  \forall t \in T \hspace{10pt} \text{(dual: } \lupdn_{r,t} )
\end{equation}
\textbf{For FO buyers:} $\forall i \in G^B, \forall t \in T \hspace{12pt} (\forall r \in R, $ or $\forall s \in S$)
\begin{equation}\label{eqn_5}
\hddn_{i,r,t} + \sdd_{i,r,t} \leq P_{i, r+1,t} - P_{i, r, t}   
\end{equation}
\begin{equation}\label{eqn_6}
\begin{split}
p^{DA}_{i,t}+ \sum_{ \substack{r \in \{1, \ldots, s-1\} \\ s \neq 1} } \left( \hddn_{i,r,t}+\sdd_{i,r,t}\right) - \\ \sum_{\substack{r \in \{s, \ldots, \lvert {S} \rvert-1\} \\ s \neq \lvert {S} \rvert} } \left(\hdup_{i,r,t}+\sdu_{i,r,t}\right)
= P_{i,s,t} 
\end{split}
\end{equation}
\begin{equation}
\label{eqn_7}
\begin{split}
\sum_{ \substack{r \in \{1, \ldots, s-1\} \\ s \neq 1} } \left( \hddn_{i,r,t}+\sdd_{i,r,t}\right) +\\  
\sum_{\substack{r \in \{s, \ldots, \lvert {S} \rvert-1\} \\ s \neq \lvert {S} \rvert} } \left(\hdup_{i,r,t}+\sdu_{i,r,t}\right) \leq y_{i,s,t}
\end{split}
\end{equation}
\begin{equation}
\label{eqn_8}
y_{i,s,t} \geq p^{DA}_{i,t} - P_{i,s,t}  
\end{equation}
\begin{equation}
\label{eqn_9}
y_{i,s,t} \geq P_{i,s,t} - p^{DA}_{i,t}  
\end{equation}
\textbf{For FO sellers:} $\forall i \in G^S, \forall t \in T $
\begin{equation}\label{eqn_10}
\sum\nolimits_{r} \hsupdn_{i, r, t} + \sum\nolimits_{a} \beta_a \cdot res_{i, a, t} \leq u_{i,t}\cdot RR_i 
\end{equation}	
\begin{equation}\label{eqn_11}
p^{DA}_{i,t} + \sum\nolimits_{r} \hsup_{i,r,t}  ++ \sum\nolimits_{a \in {AS}^{\uparrow}} res_{i,a,t}\leq
\Pmax_{i,t} \cdot u_{i,t} 
\end{equation}	
\begin{equation}\label{eqn_12}
p^{DA}_{i,t} - \sum\nolimits_{r} \hsdn_{i,r,t} - \sum\nolimits_{a \in {AS}^{\downarrow}} res_{i,a,t}  \geq \Pmin_{i,t}\cdot u_{i,t} 
\end{equation}	

In brief, the additional terms in the objective function reflect the probability-weighted RT costs of FO suppliers \eqref{eqn_1b} and buyers \eqref{eqn_1c}, and resources that self-hedge their DA schedules \eqref{eqn_1d}. Constraint \eqref{eqn_4} balances supply and demand for up and down FO.  For FO buyers, \eqref{eqn_5} limits the volume of  down FO; and \eqref{eqn_6} ensures  FO volume  is adequate to manage imbalances between the forecasted RT upper operating limit of each FO buyer $P_{i,s,t}$ and their DA energy award $p^{DA}_{i,t}$. For FO sellers, eqs. \eqref{eqn_10}-\eqref{eqn_12} ensure that the FO sold are within the seller's technical capabilities, i.e., hourly ramping \eqref{eqn_10}; generating capacity \eqref{eqn_11}; and minimum generation level \eqref{eqn_12}. Last, objective function term \eqref{eqn_1e} in conjunction with eqs. \eqref{eqn_7}-\eqref{eqn_9} help choose among alternative optima. For a more detailed discussion of this formulation and simple examples, please refer to \cite{FOpaper1}. For units that can start within an hour, we allow them to provide upward flexibility even if they are not online (i.e., $u_{i,t}=0$). To approximate their costs, we include an additional variable for them $u^{RT}_{i,r,t}$ and we do require that $u_{i,t}+\sum_r{u^{RT}_{i,r,t}} \leq 1$. The flexibility up they provide cannot exceed their minimum level and any ramping they can achieve within 60 minutes after their start-up time. Last, we include in the objective function a term to approximate their start-up costs:  $\sum_r{u^{RT}_{i,r,t} \cdot {STARTUPCOST}_i \cdot \Pi^{\uparrow}_{r}}$.

\section{Case study}
\subsection{Test System}
\begin{figure}[htb!]
\small
\begin{center}
\includegraphics[scale=1]{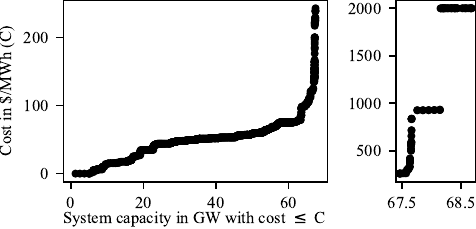} 
\caption{Supply curve excluding wind and solar with cost $\leq$ 250 \$/MWh (left) and $\geq$ 250 \$/MWh (right).}
    \label{fig:supply_curve}
\end{center}
\end{figure}

We simulate an ERCOT-like system, which is based on a synthetic dataset that has an installed capacity mix in line with ERCOT's 2019 historical data. In detail, the system includes 26 GW wind, 2 GW solar, 8 GW (175 units) that can be committed up to an hour ahead, and 61 GW (295 units) that can only be committed DA. Fig. \ref{fig:supply_curve} shows the system-wide supply curve, which has two very steep portions at the beginning and the end of it. The supply curve is based on DA energy bid curves submitted to ERCOT. 

The timeseries inputs are based on weather year 2018. For each day, 5,000 scenarios for wind, solar, and load are downloaded from \cite{ORFEUS_data}. FESTIV runs are computational expensive. Hence, we choose  twelve characteristic weeks by clustering the 51 weeks available in \cite{ORFEUS_data} based on DA and RT features.\footnote{For the DA clustering the features are the average mean and standard deviation of the DA net load. For the RT, the features are the mean and standard deviation of the normalized by the DA standard deviation difference of RT actual and DA mean net load.} For each cluster, we choose a \textit{characteristic} week that has the lowest distance from other members in the cluster. To report annual results, we multiply simulation results of each characteristic week by the number of weeks in its cluster. The characteristic weeks are scattered across seasons and reflect a wide variety of system conditions. 

\begin{figure}[htb!]
\small
\begin{center}
\includegraphics[scale=1]{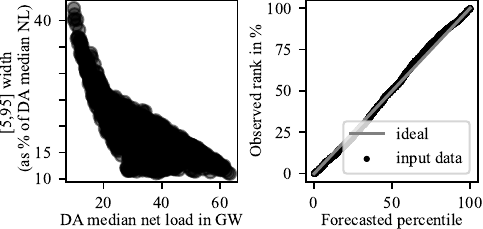} 
\caption{(Left) Width of [5,95] prediction interval as a percentage of median net load. (Right) Observed rank vs forecasted percentile of net load.}
    \label{fig:width_pplot}
\end{center}
\end{figure}

\begin{figure}[htb!]
\small
\begin{center}
\includegraphics[scale=1]{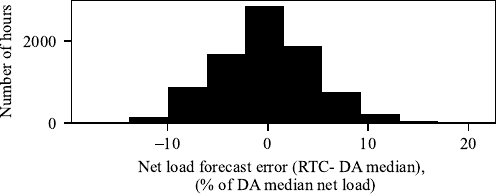} 
\caption{Net Load Forecast error }\label{fig:forecast_error}
\end{center}
\end{figure}

Assessing the forecast quality of net load,\footnote{For each of system-wide load, wind, and solar, we estimate 99 percentiles (1-99) from the 5,000 scenarios provided in \cite{ORFEUS_data}. To estimate percentiles for net load, we use the load value at the respective percentile and subtract from it the $100-percentile$ of wind and solar.} we observe that the probabilistic forecast is relatively sharp, especially for higher values of median DA net load (see left chart in  Fig. \ref{fig:width_pplot}). The probabilistic forecast is also well calibrated as shown by the proximity of the observed rank to the forecasted percentile (see right chart in Fig. \ref{fig:width_pplot}). We use the median as the \textit{deterministic} DA forecast, which seems a sensible choice as it has very low bias (mean error -0.1\%) and errors within a range of -18\% to 21\% (see Fig. \ref{fig:forecast_error}).  

\subsection{Product Parameters}
For both products (IR and FO), we must choose a set of discrete levels that reflect possible RT net load values. The levels determine the number of steps of the IR demand curve and the number of FO that can be used to hedge RT outputs of net load constituents.  As the number of constraints and variables increases with the number of levels, we choose nine levels (i.e., $|S|=9$ for FO, 4 steps for up/down IR demand curve). 

\begin{figure}[htb!]
\small
\begin{center}
\includegraphics[scale=1]{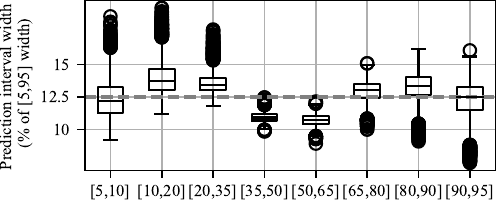} 
\caption{Relative size of FO tiers/steps in IR demand curve. Each boxplot includes the width of respective prediction intervals normalized by the width of the [5,95] prediction interval.}
    \label{fig:tiers_size}
\end{center}
\end{figure}

For this first large-system demonstration of FO and IR/FO comparison, we choose  the $5^{th},10^{th},20^{th},35^{th},50^{th},65^{th},80^{th},90^{th},95^{th}$ percentiles. As Fig. \ref{fig:tiers_size} shows, with this choice, the relative tier size is on average similar among tiers. Sometimes, the tier size is smaller at higher percentiles of  net load. This seems appropriate as the real-time supply curve  usually gets steeper as net load increases. Whereas this choice seems sensible for this article, future research could propose methodologies that choose the subset of percentiles in an optimal manner. 

For the FO, we need to choose strike-up and strike-down prices. We hypothesize that the strike-up price will be higher than the DA cost because resources might incur some last-minute costs (e.g., to procure fuel). Similarly, we hypothesize that the strike-down price will be lower than the DA cost because resources might have already incurred some costs to prepare for RT production (e.g., they might need to pay `parking' fees for unused fuel). In this article, we assume that in RT, units will offer electricity higher and lower than their DA energy schedule at the highest and lowest cost of their DA cost curve, respectively. We also assume that units do not have access to good options for self-hedging upward imbalances, which means that any scarcity of upward FO could trigger reserve scarcity in RT. That's why we choose \$225/MWh as the value for $\widetilde{\vcup_{i,t}}$, which is equal to the probability of the most infrequent upward tier multiplied by the spinning reserve scarcity price. We choose \$0/MWh as the value for $\widetilde{\vcdn_{i,t}}$, assuming that we can curtail renewable generation at zero cost. Based on our familiarity with the problem, we choose 2.8 for $M$ because this choice gives us negligible amount of up and down options that can be simultaneously exercised. This choice might harm the cost-efficiency but it reduces the volume of FO contracts managed by the system operator \cite{FOpaper1}. 

\begin{figure}[htb!]
\small
\begin{center}
\includegraphics[scale=1]{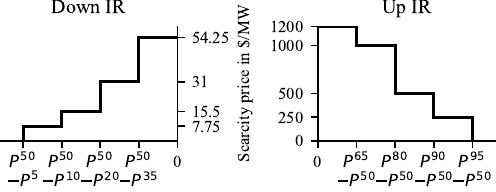} 
\caption{IR demand curves in our simulations. The upward demand curve has steps that procure reserves to cover imbalances: within the [50,65] prediction interval of  net load at a scarcity price of \$1,200/MW; within the [65,80] prediction interval of net load at a scarcity price of \$1,000/MW; and so on.}
    \label{fig:IRdemand}
\end{center}
\end{figure}

For IR, we choose scarcity prices based on earlier CAISO proposals (see Fig. \ref{fig:IRdemand}). Note though that the latest CAISO design includes a flat scarcity price of \$55/MW \cite{DAME_Final}. 

\section{Results}
We report the results for each party that would be affected by the introduction of the two products, starting with the system operator, continuing with suppliers of flexibility ($G^S$), and concluding with resources that introduce imbalances ($G^B$). 

\subsection{System Operator}
The system operator aims to facilitate cost-effective and reliable power system operations and to ensure just, reasonable, and not unduly discriminatory prices while remaining revenue adequate. 

\subsubsection{System Costs}
In IR simulations, the system cost is slightly lower compared to the FO simulations, but the difference falls within the DA MIP gap (0.5\%). Therefore, the simulated system costs, reported in Table \ref{table: costs}, suggest that both products are equally effective in supporting cost-effective and reliable operations for the study period.  

\begin{table} [htb!]
\tiny
\begin{center} 
\caption{Annual System Costs for IR and FO simulations in \$ million}
\label{table: costs}
\begin{tabular}{|c|c|c|c|c|} 
 \hline
 & DA costs & RTD costs & RTD scarcity penalties & System cost \\ \hline
IR & 8,152.9 &  446.6 & 6.9 & 8,606.4 \\ \hline
FO & 8,233.4  & 391.5 & 2.6 & 8,627.4\\ \hline
\end{tabular}
\end{center}
\end{table}

\begin{figure}[htb!]
\small
\begin{center}
\includegraphics[scale=1]{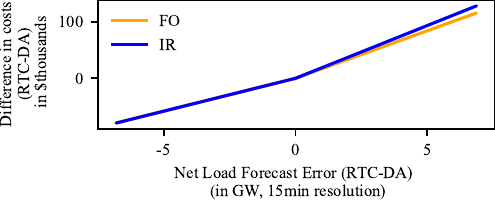} 
\caption{Linear cost curves for IR and FO simulations, estimated based on simulation outputs (see footnote 7). }
    \label{fig:linear_supply_curves}

\end{center}
\end{figure}

Whereas the total system cost is similar in both sets of simulations, its composition differs. With FO, the DA costs are higher, while the RT costs and penalties for insufficient RT reserves are lower. The simulations show that using FO can reduce RT costs, especially in cases of upward RT imbalances (i.e., when RT net load is higher than DA net load). As Fig. \ref{fig:linear_supply_curves} shows,\footnote{ For each 15-min simulated interval, we record the net load forecast error and the RTC production costs. We obtain the difference in production costs by subtracting the DA production costs from the RTC costs for the same time interval. For each set of simulations (FO, IR), we perform two linear regressions, which have the net load forecast error as the independent variable and the difference in production cost (RTC-DA) as the dependent variable. One linear regression is performed for the subset of intervals with positive net load forecast errors and another one for the subset of intervals with negative net load forecast errors. In FO simulations, R$^2$ is 0.82 and 0.85 for positive and negative errors, respectively; in  IR simulations, R$^2$ is  0.87 and 0.84 for positive and negative errors, respectively.} RT cost curves are less steep for FO vs. IR, particularly in the upward direction. The difference in the slope is explained by different mix of flexibility suppliers, as we later discuss in section \ref{section: flexsupp}. 

Our simulations adhere to standard practices, evaluating the effectiveness of a new product over a historical study period. This approach is an out-of-sample analysis, as historical errors are unknown when forecasts are generated.\footnote{Unfortunately, this is not entirely true here as \cite{Carmona_Xinshuo} had to use some future historical realizations to create the wind forecasts.} However, our sample size is small because it only simulates one scenario (the historical) per day, and net load forecast errors are known to exhibit high autocorrelation. To gauge the statistical robustness of our results, we conduct simulations for all 5,000 scenarios provided in \cite{ORFEUS_data} using a simple model for RT, provided in Appendix A. Fig. \ref{fig:simple_model_stats} shows the weekly difference between IR and FO system costs for the 12 weeks simulated.

\begin{figure}[htb!]
\small
\begin{center}
\includegraphics[scale=1]{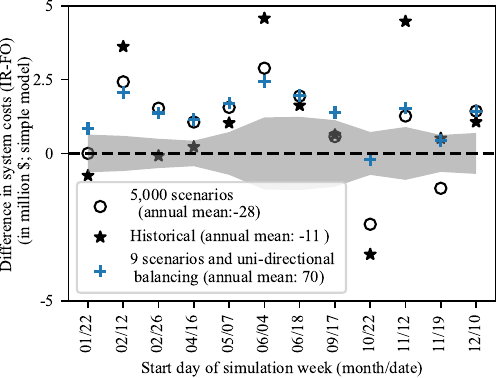} 
\caption{Difference in weekly system costs (IR-FO) simulations, as calculated by the simple model in Appendix A. The shaded area shows the MIP gap (i.e., $\pm 0.5\%$ of DA costs).}
    \label{fig:simple_model_stats}
\end{center}
\end{figure}

The analysis over the 5,000 scenarios also finds that the annual system costs are slightly lower in the IR vs. the FO simulations. However, the weekly differences in system costs are relatively high. For 10 out of 12 weeks, the differences are larger than the MIP gap (e.g., any week other than weeks starting on Jan 22 and Sep 17). In this case, we would expect system costs to be lower in FO compared to IR simulations, i.e., the circles are above the shaded area in Fig. \ref{fig:simple_model_stats}. This is indeed the case for 80\% of the weeks with substantial differences in costs between the two sets of simulations. 

However, for two weeks (week starting on Oct 22 and week starting on Nov 19), the IR simulations have lower system costs than FO simulations (i.e., circles below the shaded area of Fig. \ref{fig:simple_model_stats}). At first glance, this result contradicts our hypothesis that the model with FO can choose a DA schedule with lower or equal probability-weighted system costs than the model with IR. However, due to practical limitations, the FO model considers only an approximate 9-point discrete distribution of the 5000 scenarios and a subset of balancing actions (e.g., only increases in production when RT net load is higher than DA net load). Due to these practical limitations, the FO model has a narrow view of RT operations and identifies DA schedules that are beneficial within this narrow view but are not necessarily optimal when the full spectrum of RT scenarios and recourse actions is considered.  

In conclusion, the out-of-sample analysis indicates that the annual results are statistically robust. The analysis of weekly results suggests that further research on selection of the discrete levels  ($S$) and the parameter that controls the FO volumes ($M$)  can result in increased benefits.

\subsubsection{Revenue Adequacy}
To ensure just and fair prices, system operators follow cost-causation principles, which means that charges are approximately in line with the benefits participants receive or the costs they introduce. Ideally, by following these principles, the system operator would recover all the costs associated with a product they have determined the demand for. However, our simulations suggest that the system operator is not revenue-neutral with IR. It only recovers about 50\% of the DA IR procurement costs (Table \ref{table: rev-neu}, row 1: 25.93/51.96) from resources with RT imbalances and must find another way to recover the remaining 50\%, possibly through a uniform charge among all participants.

\begin{table} [htb!]
\tiny
\begin{center} 
\caption{Product-related Settlements in \$ million }
\label{table: rev-neu}
\begin{tabular}{|c|c|c|c|} 
 \hline
 & DA $G^S/G^B$   & RT $G^S/G^B$ & ISO \\ \hline
IR & 51.96/0  & 0/-25.93 & \textbf{-26.03} \\ \hline
FO & 151.17/-151.17 & -58.76/58.76 & 0\\ \hline
\end{tabular}
\end{center}
\end{table}

The system operator recovers 100\% of the FO DA and RT costs. The FO settlement scheme is revenue-adequate because it allocates costs at each stage (DA,RT) separately. That way, the FO scheme does not suffer from errors in the forecasted distribution of imbalances. In contrast, in the IR scheme the system operator implicitly takes responsibility for divergence in the forecasted and realized distribution of imbalances. More importantly, even if the distribution of imbalances is accurately forecasted, using the DA IR price as the maximum charge to participants with imbalances leaves the system operator responsible for part of the costs that cannot be recovered due to the difference between the average observed imbalance and the high percentile of imbalance, used to determine the reserve requirements.


\subsubsection{Flexibility Price Signals}
\begin{figure}[htb!]
\small
\begin{center}
\includegraphics[scale=1]{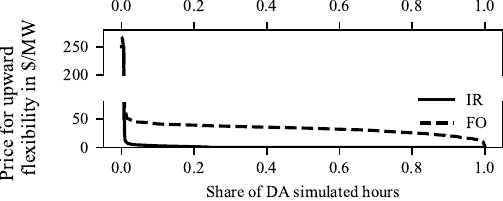} 
\caption{Price duration curve for IR/FO up price. \textit{Note:} For FO up options, each tier has a different price. We here show the highest price among all tiers with positive volume of contracts. }
\label{fig:flexup_price_duration_curve}

\end{center}
\end{figure}

The DA price signals for IR and FO differ significantly. The FO price is higher and more dynamic, as shown in Fig. \ref{fig:flexup_price_duration_curve}. The price is higher because the FO suppliers agree to give up any profit margins exceeding their stike price. The price varies based on system conditions, which highlights how important is to price FO within the DA market, as discovering these prices through private contracts would be very difficult.

\subsection{Suppliers of Flexibility} \label{section: flexsupp}
Results show that IR and FO simulations schedule different types of flexible suppliers and reward them differently.

\subsubsection{Supplier Mix}
\begin{figure}[htb!]
\small
\begin{center}
\includegraphics[scale=1]{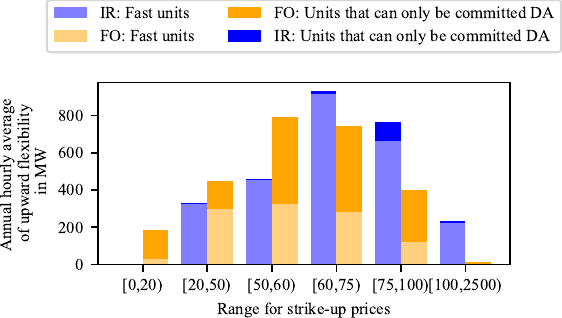} 
\caption{ Supply mix for upward flexibility in IR and FO simulations. Suppliers are grouped based on strike-up prices. }
    \label{fig:Up_supply_mix}
\end{center}
\end{figure}

\begin{figure}[htb!]
\small
\begin{center}
\includegraphics[scale=1]{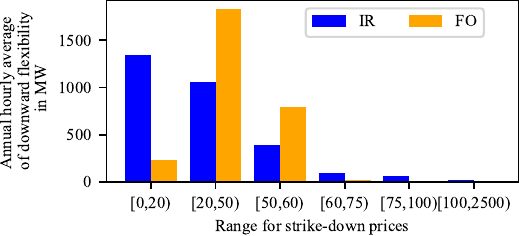} 
\caption{ Supply mix for downward flexibility in IR and FO simulations. Suppliers are grouped based on strike-up prices.}
    \label{fig:Dn_supply_mix}
\end{center}
\end{figure}

Compared to IR simulations, the FO simulations procure upward and downward flexibility from suppliers with lower and higher strike prices, respectively (see Figs. \ref{fig:Up_supply_mix} and \ref{fig:Dn_supply_mix}). This result is expected due to the modified DA objective function in FO, which considers probability-weighted RT cost increases and savings to manage imbalances. A substantial portion of the less expensive upward flexibility in FO comes from units that can only be committed DA. In fact, 59\% of the total FO up volume is provided by units that can only be committed in DA and the respective percentage is only 5\% for IR up. 

\begin{figure}[htb!]
\small
\begin{center}
\includegraphics[scale=1]{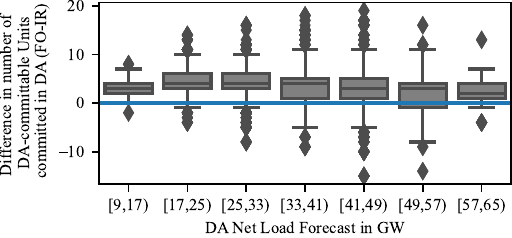}
\caption{Difference in number of units that can only be committed in DA and they are indeed committed in FO and IR simulations.}
    \label{fig:stats_units}
\end{center}
\end{figure}

Both IR and FO have the opportunity to commit units in DA that could provide flexibility in RT. However, the IR simulations only acknowledge the downside of procuring flexibility from additional DA-committable units, which is additional commitment costs. In contrast, the FO simulations recognize in addition to the downside, the upside: access to flexibility, which is potentially less expensive flexibility than the one that can be offered by fast units in RT. In this case study, the FO commits more units that can only be committed in DA 90\% of the time. As shown in Fig. \ref{fig:stats_units}, the FO simulations commit more of those units, especially during the hours that the forecasted net load is relatively low (between 17 and 35 GW). This might be explained by the relatively steeper portion of the cost curve at the left tail of the supply curve (see Fig. \ref{fig:supply_curve}) and the plethora of alternative feasible schedules for hours with lower net loads. 

\subsubsection{Financial Incentives}
The suppliers of flexibility (IR, FO) receive multiple revenue streams to manage imbalances caused by forecast errors. In DA, they receive payments for reserving capacity for IR and FO. In RT, they receive or pay RT energy prices for increasing or decreasing their RT output from their DA level, thereby increasing their revenue or reducing their costs by effectively outsourcing their DA obligations. Lastly, in the case of the FO, suppliers of flexibility payoff the option buyers.  We summarize daily statistics for all those cashflows in Table \ref{table: flexpay}.  


\begin{table} [htb!]
\tiny
\begin{center} 
\caption{Daily Cashflows for Flexible Participants ($G^S$) in \$ million }
\label{table: flexpay}
\begin{tabular}{|E|E|E|E|E|E|} 
 \hline
 & 1:DA up/down flexibility & 2: RT energy & 3: RT up/down & 4: Flexibility Revenues (1+2+3) & 5: Margins (4 - RTC costs)\\ \hline
IR (mean) & 0.14/0 &  0.29 & N/A & 0.43 & 0.18\\ \hline
FO (mean) & 0.42/0.07  & 0.07 & -0.16/-0.04 & 0.37 & 0.31\\ \hline
IR (std) & 0.16 &  1.64 & N/A & 1.63 & 1.35 \\ \hline
FO (std) & 0.28/0.03  & 1.22 & 0.15/0.04 & 1.11 &0.82\\ \hline
\end{tabular}
\end{center}
\begin{center}
\vspace{-1em}
      \tiny \item Table note: We report results as if all resources in $G^S$ are under a single account. Please refer to Appendix B for details on the calculations.
\vspace{-1em}
\end{center}
\end{table}

The payments to flexible suppliers in DA are three times higher in FO compared to IR (0.42 vs 0.14 in Table \ref{table: flexpay}). The higher DA revenue for FO is reasonable because the FO suppliers have to pay-off the FO buyers in RT. In this case study, the flexibility-related revenues are slightly higher (+16\%) in IR compared to FO, but they are much more volatile (the coefficient of variation is 3.8 and 3 in IR and FO, respectively). Despite the higher level of flexibility-related revenues in IR, the profit margins are lower on average in IR due to the higher re-dispatch costs. In addition to lower mean daily values, the IR margins also appear more volatile than the FO margins, with higher standard deviations.  

Thus far, the analysis of the case study suggests that flexible suppliers would favor the FO market due to the higher average margins and lower volatility compared to the IR market. However, one potential drawback of the FO market from a flexible supplier's standpoint is its exposure to the RT price. In this case study, we observe that several units had sold FO that were activated (e.g., strike up price lower than RT price and upward imbalance), but the optimization engine of the RT market scheduled them at different levels than the ones implied by the activated FO volume (i.e., RT schedule $<$ DA schedule + FO up activated). This issue merits further investigation in future research, with alternative assumptions on RT bidding or use of agent-based model simulations.   

\subsection{Resources introducing DA-RT Imbalances}
Contrasting the simulation results from IR and FO, we notice two significant differences for resources with imbalances caused by forecast errors. First, lower volume of flexibility is demanded in FO simulations. Second, the balancing charges are less volatile in FO simulations. 

\subsubsection{Demand for Flexibility} In IR simulations, the demand for flexibility is determined through an exogenous demand curve, while in FO simulations it is calculated endogenously. Both IR and FO consider uncertainty within the [5,95] prediction interval of wind, load, and solar. However, the procured flexibility is lower in FO compared to IR 15\% of the time, and similar for the remaining 85\% of the time. 

\begin{figure}[htb!]
\small
\begin{center}
\includegraphics[scale=1]{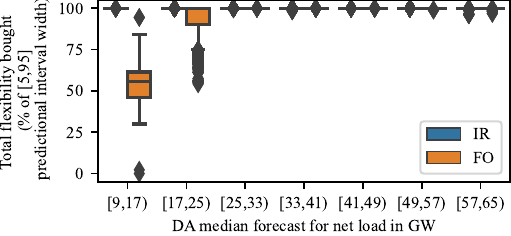} 
\caption{Demand for flexibility (in up and down directions) normalized by the [5,95] prediction interval width.}
    \label{fig:Flexibility_demand}
\end{center}
\end{figure}

Based on the data in Fig. \ref{fig:Flexibility_demand}, it is evident that the demand for flexibility in FO is lower than the width of the [5,95] prediction interval when the net load is relatively low. Specifically, in 98\% of the hours with DA median net load less than 20 GW the demand for flexibility is lower in FO vs IR. In annual terms, 13\% of hours have both DA net load lower than 20 GW and lower volume for flexibility options  than imbalance reserves. We will refer to these hours as `low net load-low FO' hours. The lower volume of the FO is mainly due to lower demand for flexibility that would enable running the system in RT at the 5$^{th}$ percentile of forecasted net load. More than 99\% and 68\% of the `missing volume' can be attributed to lower demand for down flexibility in 92\% and 100\% of the 'low net load-low FO' hours, respectively.

\begin{figure}[htb!]
\small
\begin{center}
\includegraphics[scale=1]{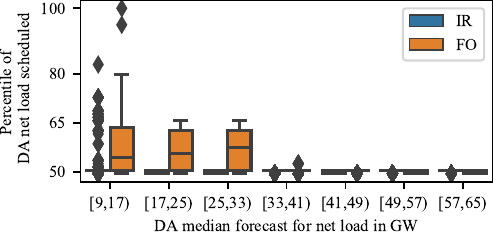} 
\caption{Forecasted percentile of DA net load schedule}
    \label{fig:percentile DA net load}
\end{center}
\end{figure}

Whereas FO might not always procure enough down flexibility to run the system at the $5^{th}$ percentile of net load, it is worth pointing out that the need for down flexibility is higher in FO than in IR 35\% of the time. This is because the FO yields DA net load schedules higher than the 50th percentile 36\% of the time, whereas the IR does the same only  1\% of the time (see Fig. \ref{fig:percentile DA net load}). 

This higher need for down flexibility is met for net load higher than 20 GW. Therefore, we can conclude that sometimes the FO finds beneficial to request more dowwnward flexibility volume than the width of the $[5,50]$ prediction interval. This outcome could be the result of asymmetric up and down balancing costs and illustrates the interplay between the demand for flexibility and the DA schedule of uncertain resources.

\subsubsection{Balancing Charges}
Uncertain resources always have to settle their imbalances in the RT energy market. In addition to the RT energy market cashflows, uncertain resources have to pay for IR and FO in RT and DA, respectively. Under FO, uncertain resources also receive FO payoffs in RT. We summarize all daily cashflows in Table \ref{table: uncpay}. 
The data in Table \ref{table: uncpay} indicate that the  balancing charges of the IR scheme are both higher on average and  more volatile compared to FO (see last column in Table \ref{table: uncpay}). 

\begin{table} [htb!]
\tiny
\begin{center} 
\caption{Daily Imbalance-related Cashflows in \$ million }
\label{table: uncpay}
\begin{tabular}{|E|E|E|E|E|} 
 \hline
 & DA up/down  & RT energy & RT up/down & Total Payment  \\ \hline
IR (mean) & N/A &  -0.29 & -0.29 & -0.58 \\ \hline
FO (mean) & -0.42/-0.07  & -0.07 & 0.16/0.04 & -0.37 \\ \hline
IR (std) & N/A &  1.64 & 0.93 & 2.23  \\ \hline
FO (std) & 0.28/0.03  & 1.22 & 0.15/0.04 & 1.11 \\ \hline
\end{tabular}
\end{center}
\begin{center}
\vspace{-1em}
      \tiny \item Table note: We report results as if all resources in $G^B$ are under a single account. Please refer to Appendix B for details on the calculations.
\vspace{-1em}
\end{center}
\end{table}

Although the charges related to imbalances are lower on average under FO, it is worth noting that the net payments for DA energy (load payments -renewable revenues) are higher in FO compared to IR. However, the daily DA energy costs are 0.13 million \$ higher in FO compared to IR. This difference is smaller than the difference between 0.58 and 0.37 million \$ in Table \ref{table: uncpay}. Therefore, the results suggest that participants affected by imbalances won't be unfairly harmed by being held responsible for DA FO premiums.

\section{Conclusions}
We perform simulations with two alternative products for management of day-ahead-to-real-time imbalances caused by forecast errors: a reserve product (imbalance reserves) and a newly proposed product (flexibility options). FO have physical attributes of a reserve product and financial attributes of financial options. The FO demand is endogenously determined in the DA market. 

Our simulations for 12 characteristic weeks of a 2019 ERCOT-like system suggest that both products are almost equally effective in managing imbalances. However, the FO product yields less steep RT cost curves. The less steep RT cost curves reduce the volatility of flexibility-related revenues of flexible suppliers and of imbalance-related charges for resources subject to forecast errors. That way, the FO product addresses risk management needs of participants in a better way than IR. Moreover, the FO settlement scheme is revenue-adequate as it does not allocate costs for products that manage imbalance risk outside the market. 

The results highlight how alternative product designs can have substantially different consequences for participants' risk. Therefore, when assessing market reforms, market designers need to look beyond traditional metrics of cost-efficiency, reliability, and financial transfers into the impact of alternative market designs on risk perceived by market participants.  

Future work on this specific set of products could quantify the impact of the two products on individual resources instead of portfolio of resources; include storage as a flexible supplier; add deliverability constraints; and extend the analysis using alternative FO pricing schemes and advanced methods for parameter tuning.

\section*{Appendix A: Simple Model for Extensive Out-of-sample Analysis}

To assess the statistical robustness of our results, we perform RT simulations with a simpler than FESTIV model over 5,000 scenarios. The model increases (${p}^{RT,+}_{i,t}$) or reduces (${p}^{RT,-}_{i,t}$) electricity generation by IR/FO suppliers to minimize balancing costs. Whereas FESTIV optimizes the commitment of all RT-committable units in RT, the simple model turns on FO/IR suppliers only. FESTIV re-optimizes ancillary service schedules in RT, whereas the simple model considers them fixed. The simple model is as follows:
\begin{equation}
\small
\label{optprobsimple}
		\begin{aligned}
		\underset{\Xi}{\text{minimize}}  \sum_{i,t}({\ccup_{i} \cdot {p}^{RT,+}_{i,t} - \ccdn_{i} \cdot {p}^{RT,-}_{i,t}} )\\ +\sum_t({\overline{\lambda} \cdot\epsilon_t^{\uparrow}-\underline{\lambda} \cdot \epsilon_t^{\downarrow}}) + \sum_{i,t} {C^{No load}_i \cdot u^{RT}_{i,t} } + \\\sum_{i,t} {C^{Start up}_i \cdot (u^{RT,start}_{i,t} -u^{DA,start}_{i,t} )}\\
\text{ where } \Xi = \{ {p}^{RT,+}, {p}^{RT,-} ,\epsilon_t^{\uparrow},\epsilon_t^{\downarrow},u^{RT}_{i,t},u^{RT,start}_{i,t} \}
		\end{aligned}
\end{equation}

\begin{equation} \label{up_simple}
{p}^{RT,+}_{i,t} \leq \sum\nolimits_r{{HS}^{\uparrow}_{i,r,t}} \cdot (U^{DA}_{i,t} +u^{RT}_{i,t})
\end{equation}
\begin{equation} \label{down_simple}
{p}^{RT,-}_{i,t} \leq \sum\nolimits_r{\hsdn_{i,r,t}}
\end{equation}
\begin{equation} \label{start_simple}
u^{RT,start}_{i,t} \geq (U^{DA}_{i,t} +u^{RT}_{i,t}) -(U^{DA}_{i,t-1} +u^{RT}_{i,t-1})
\end{equation}
\begin{equation} \label{balance_simple}
\sum\nolimits_i({p}^{RT,+}_{i,t}-{p}^{RT,-}_{i,t}) +\epsilon_t^{\uparrow} -\epsilon_t^{\downarrow} = \widetilde{{NL}^{DA}} - {NL}_{t}^{DA schedule}
\end{equation}

In brief, eqs. \eqref{up_simple} and \eqref{down_simple} allow the output of a unit $i$ to increase or reduce according to the FO/IR awards. Eq. \eqref{start_simple} keeps track of unit start-ups. Last, eq, \eqref{balance_simple} balances the net load forecast error with upward and downward movements.

\section*{Appendix B: Metrics}
We calculate RT energy payments by  uncertain participants (which is equal to RT energy revenue for flexibile participants) per the following formula: $\sum_{i \in G^B}{\lambda^{RTC}_t \cdot (p^{RT}_{i,t}-p^{DA}_{i,t})}$. For the remaining metrics in FO, we use formulas from \cite{FOpaper1}. We calculate the RT need for FO volume as if load, wind, and solar are managed under a single account. For IR, we calculate the observed imbalances separately for load, wind, and solar to allocate the IR procurement costs.

\bibliography{IEEEabrv,Bibliography}
\bibliographystyle{IEEEtran}

\end{document}